\documentclass[preprint,12pt]{elsarticle}

\usepackage{graphicx}
\usepackage{amsmath}
\usepackage{amssymb}
\usepackage{amsthm}
\usepackage{booktabs}
\usepackage[update,prepend]{epstopdf}
\usepackage[slantedGreek]{mathptmx}
\usepackage{url}
\usepackage{bbm}
\usepackage{accents}
\usepackage{subfigure}

\newlength{\dhatheight}

\renewcommand{\div}{\mbox{\rm div}}

\newcommand{\xb}{\bar{x}}
\newcommand{\ab}{\bar{m}}

\newtheorem*{theorem*}{Theorem}

\journal{Journal Mathematical Biology}

\begin{document}

\begin{frontmatter}

\title{ Adaptive Learning in Large Populations}
\author{Misha Perepelitsa }

\date{\today}
\address{
misha@math.uh.edu\\
Department of Mathematics\\
University of Houston\\
4800 Calhoun Rd. \\
Houston, TX.}

\begin{abstract}

We consider the adaptive learning rule of Harley (1981) for behavior selection
in symmetric conflict games in large populations. This rule uses  organisms' past, accumulated rewards as the predictor for future behavior, and  can be traced in many life forms from bacteria to humans. We derive a partial differential equation (PDE) that describes the stochastic learning in a heterogeneous population of agents. The equation has a structure of the conservation of mass type equation  in the space of stimuli to engage in a  particular behavior. We analyze the solutions of the PDE model for symmetric 2x2 games. It is found  that in games with small residual stimuli, adaptive learning rules with larger  memory factor converge faster to the optimal outcome.

\end{abstract}

\begin{keyword}
Adaptive learning \sep relative payoff sum \sep symmetric games



\end{keyword}

\end{frontmatter}

\begin{section}{Introduction}

%Natural selection is a mechanism for selecting a fittest behavioral trait in a population of organisms.

The seminal paper Maynard Smith \& Price (1973) introduced the concepts of game theory into the study of animal behavior to explain the evolution of  behavioral traits. To this end,  a notion of evolutionarily stable strategy (ESS) was developed to describe stable outcomes of natural selection, by defining it as being  uninvadable by pure or mixed strategies players.   A dynamical process leading to an ESS can be formalized by the following model. Consider a situation where each individual in a population  consistently uses  one  of the available behavioral traits $t_1,..,t_k$ to interact with other members and no mutations take place. Assume the  choice of the behavior (action) is inherited, and the population consists of groups that use a particular behavior.  Over a long period of time involving large number of interactions, the average fitness per game for individuals  using $t_i,$ denoted by $W_i,$ is compared to the population average fitness $\bar{W}$, and the individuals in $t_i$-group are  reproduced at rate proportional to $W_i/\bar{W}.$ If the changes in the frequencies of $t_i$ are approximately continuous  Taylor \& Joniker (1978),  Zeeman (1981) derived the replicator dynamics equations for the frequencies  $m_i$ of individuals acting according to  $t_i:$
\begin{equation}
\label{intro:rep}
\frac{dm_i}{dt}{}={}km_i(W_i-\bar{W}),\quad i=1..N
\end{equation}
and showed that ESS's are the asymptotically stable fixed points of this system of equations.

Considered from the point of view of  game theory, an evolutionarily stable strategy is  a refinement of a Nash equilibrium, which describes an optimal choice of actions in games. This way, natural selection is a mechanism for implementing a rational decision-making in the evolution of species.  There is another way by which organisms, even without complex cognition, can discover optimal actions.  It can be achieved through their ability to regulate behaviors depending on the  experience, in particular, through the tendency to repeat positive and avoid negative experiences. This is known as the law of effect, first formulated by Thorndike (1989), and generally accepted as one of the main paradigms of animal behavior, see for example,  Ferster \& Skinner (1957), Herrnstein (1961),  Herrnstein (1970),  Catania (1963), Chung \& Herrnstein (1967), Domjan \& Burkhard (1986).

The law of effect is the basis for the reinforcement learning models. They were introduced by Bush \& Mosteller (1955), and since then have been applied  to problems in such diverse areas as  biology, economics, and engineering. Some  representative examples of the extensive literature on this subject can be found in Harley (1981), Cross (1983), Roth \& Erev (1995), Erev \& Roth (1998), Sutton \& Barto (1998), Sandholdm (2010), Nax \& Perc (2015).

Interestingly enough, Börgers \& Sarin (1997) demonstrated that models of learning also lead to the replicator dynamics equations similar to \eqref{intro:rep}. For that, they considered repeated plays of a 2x2 game between two agents who adjust their probabilities for actions according to a reinforcement learning model of Cross (1983), and derived the replicator dynamics equation in the limit of small payoffs. More general situations, both in terms of the type of the games, the kind of reinforcement models and their relation to replicator-like equations are discussed in a monograph by Fudenberg \& Levine (1998), Rustichini (1999),  Fudenberg \& Takahashi (2011), and Mertikopoulos \& Sandholm (2016).

The law of effect can be expressed in many different ways, depending on which decision-making facilities are reinforced and on the specific rules of reinforcement. In the model of Cross (1983), it is the probability to play particular action that undergoes reinforcement.  In an alternative model motivated by bio-chemical processes in neural circuits, Harley (1981) proposed the relative payoff sum (RPS) algorithm as a decision making mechanism. The RPS learning rule assumes the ability of organism to maintain a record of cumulative rewards from previous experiences, which at epoch $t$ is given by vector  $S(t)=(S_1(t),..,S_k(t)).$ $S(t)$ is the predictor for future behavior and can be interpreted as a vector of 
``motivations'' or ``stimuli'' to engage in a corresponding action.  From the current stimuli an agent computes the probability to play $t_i:$  
\[
\frac{S_i(t)}{\sum_j S_j(t)}.
\] 
If action $t_{i_0}$ is chosen and 
it brings payoff $P_{i_0}(t),$ which is assumed to be non-negative, then stimuli are updated according to the rule
\begin{equation}
\label{RPS1}
S_{i_0}(t+1){}={}(1-\bar{\mu})S_{i_0}(t) + \bar{\mu} r_{i_0} + P_{i_0}(t),
\end{equation}
and for $k\not=i_0,$
\begin{equation}
\label{RPS2}
S_{k}(t+1){}={}(1-\bar{\mu})S_{k}(t) + \bar{\mu} r_{k}.
\end{equation}
Positive parameter $\bar{\mu}$ expresses memory effect: payoff $P_i(t-n)$ from $n$ previous plays
will appear with the weight $(1-\bar{\mu})^{n+1}$ in the expression for $S_i(t+1).$ Parameter $r_i$ is some default level (residual) of stimuli $S_i.$ For example, the residuals might represent the genetic preference for the action. A similar learning model was introduced by Roth \& Erev (1995).

%The question then arises  which strategy, if any, will be learned by organisms implementing RPS rule.  Harley (1981) presented numerical simulations of learning in several classical   games that showed that RPS in general leads to the discovery of the optimal strategies.

The goal of this paper is to address the question of behavior of RPS learning agents in large populations, where agents are randomly matched in pairwise encounters, i.e. learning in heterogeneous populations. Unlike the evolutionary dynamics framework, RPS learning agents can not be identified with some particular strategy they  use all the time.  The RPS rule will in general prescribe new strategy every time an agent plays.

 In these situations, the natural quantity to describe the state of agents is the distribution of agents according to their current stimuli.  In large populations, the probability density function of this distribution can be approximated by  a continuous function and its changes can be described by a non-linear Fokker-Planck  equation. The derivation of this equation, which generalizes the replicator dynamics equation for heterogeneous populations, is  contained in the Appendix of this paper. The assumptions needed for the derivation of the equation are:  large population size, large  number of plays of the game,  and incremental (infinitesimal) structure payoffs. This approach is a well-known method for modeling multi-agent systems in the problems of physics, biology, economics and sociology, see for example, Risken (1992) and  Pareschi \& Toscani (2014). A similar method has been used by Traulsen,  Claussen  \& Hauert  (2005, 2006), in the analysis of the evolutionary selection by Moran process.

After deriving the equations we apply them to determine the behavior of RPS learners in symmetric 2x2 games. It should be kept in mind, however, that the time asymptotic 
behavior based on the Fokker-Planck equation and the time asymptotic of the original discrete-time stochastic process
are not, in general, the same. Moreover, the PDE model derived in this paper is a leading approximation of a continuous-time stochastic process. Thus, any statement claimed in  this paper about the convergence of the system to a particular state should be understood as a statement that the system gets close to this state within the limits of the validity of the PDE approximation. 

In game theory, the method of stochastic approximation by Benaïm \& Hirsch (1999) is typically used to determine the long-time behavior interacting agents. The method relies on the stability analysis of an averaged (deterministic) system of ODEs. As the dimension of a system is of the order (the number of agents ) $\times$ (the number of strategies per agent), even for small populations it is an extremely challenging task.  The Fokker-Planck equation, on the other hand, is applicable for large populations, the trade-off being the  loss of information about  a particular player.

The analysis of the PDE model derived in this paper shows that for games with a single Nash equilibrium,
 the strategies of all agents converge to the dominant strategy, when the RPS rule has no memory factor, or with a memory factor and zero residuals. For learning models with a memory factor the convergence is faster than for the models with perfect memory. Additionally, the learning time in the former case varies inversely with the size of the memory factor. For games with a mixed Nash equilibrium, the learning process converges to a state in which the population mean probability equals to the equilibrium value. As the population mean probability approaches its limit, the ``strength'' of the learning decreases, and  the individual probabilities do not, in general, converge to the equilibrium value. In this case, the population remains heterogeneous.  Finally, if the memory factor is present and residuals are not zeros, agent's strategies converge to some mixed strategy, for a generic 2x2 symmetric game.

\end{section}

\begin{section}{The Model}
\label{sec1}
We consider a series of plays of a game between randomly selected individuals in a large population. The payoff matrix of the game is given in the table \ref{table:inc_pay}.
The behaviors are labeled A and B.
\begin{table}
\centering
\begin{tabular}{@{}lcc@{}}
\toprule
   &  A     & B  \\
A  & (ah, ah)   & (dh,\,ch)   \\
B &  (ch, dh)  &  (bh,\, bh)  \\
\bottomrule
\end{tabular}

\vspace{15pt}

\caption{The incremental payoff matrix for the game. Parameters $a,b,c,d,h$ are positive.
 \label{table:inc_pay}}

\end{table}
We will analyze the RPS learning in the games that a) have a single pure Nash equilibrium with $a>c,$ $d>b$ (or  $a<c,$ $d<b$ ); b) two Nash equilibria with $a>c,$ $b>d$; and c) mixed Nash equilibrium when $a<c,$ $d>b.$ In the latter case, there are also two asymmetric Nash equilibria $(A,B)$ and $(B,A).$

Now, let there be a group of $N$ individuals where each individual is  characterized by vector  $X^t_i{}={}(S^t_{i,1},S^t_{i,2})$ representing the accumulated stimuli to play A and B, respectively, at epoch $t.$ Suppose that two agents $i$ and $j$ are selected at random to play the game.
Agents play with the probabilities to cooperate $S^t_{1,i}/(S^t_{1,i}+S^t_{2,i})$ and $S^t_{1,j}/(S^t_{1,j}+S^t_{2,j}),$ for agent $i$ and agent $j,$ respectively. Denote the outcomes (A,A), (A,B), (B,A), (A,A), where the first is the action chosen by agent $i.$  As the result of the interaction agent $i$  increments his/her states according to the rule
\begin{equation}
\label{eq:RPS}
X^{t+\delta}_i{}={}\left\{
\begin{array}{ll}
(r_1\mu h{}+{}(1-\mu h)S^t_{1,i} + ah,\, r_2\mu h + (1-\mu h))S^t_{2,i}), & \, \rm(A,A)\\
(r_1\mu h{}+{}(1-\mu h)S^t_{1,i} + ch,\, r_2\mu h + (1-\mu h))S^t_{2,i}),  & \,\rm(A,B)\\
(r_1\mu h + (1-\mu h))S^t_{1,i},\, r_2\mu h{}+{}(1-\mu h)S^t_{2,i} + dh), & \,\rm(B,A)\\
(r_1\mu h + (1-\mu h))S^t_{1,i},\, r_2\mu h{}+{}(1-\mu h)S^t_{2,i} + bh),  &  \,\rm(B,B)
\end{array}
\right.
\end{equation}
and symmetrically for agent $j.$ $(r_1,r_2)$ are the residual stimuli and $\mu\geq0$ the parameter of the fading memory. The former is related to the memory factor $\bar{\mu}$ from the introduction by formula $\bar{\mu} = \mu h.$ Then, time moves to the next epoch $t+\delta$ and the process is reiterated. At time $t=0$ agents have generally different, initial levels of stimuli to cooperate and defect.

\subsection{Stimuli-space}

Our main interest is in the distribution  of agents in the stimuli-space $x=(s_1,\,s_2),$ $s_1,\,s_2\geq0,$ described by the density function (PDF) $f(x,t).$  In this space the straight lines through the origin represent the sets of stimuli of constant probability to cooperate when agent is playing a mixed strategy. The probability $m$ related to the slope $k$ as
$
m{}={}(1+k)^{-1}
$
so that the stimuli with preference for C are located closer to the $s_1$-axis.
For any subset $\Omega$ in the stimuli space, $\int_\Omega f(x,t)\,dx$ represents the proportion of agents with their stimuli in the set $\Omega$ at time $t.$ 

The following equation is found to be a leading order approximation for the process,
 when parameters  $h,\delta$ are small and the number of players is large.
 \begin{equation}
\label{eq:main}
\frac{\partial f}{\partial t}{}+{}\div (u(x,t)f){}={}0,
\end{equation}
and velocity $u$ 

\begin{equation}
\label{velocity}
u(x,t){}={}\frac{1}{s_1+s_2}\left[
\begin{array}{l}
\left(a\ab(t){}+{}d(1-\ab(t)\right)s_1 + \mu(r_1-s_1) ( s_1+s_2) \\
\left( c\ab(t){}+{}b(1-\ab(t)\right)s_2 +\mu(r_2-s_2)  (s_1 +s_2)
\end{array}
\right],
\end{equation}
where
\[
\ab(t){}={}\int\frac{s_1}{s_1+s_2}f(x,t)\,dx.
\]
The analysis of the equation can be understood from the behavior of the system of ODEs:
\begin{equation}
\label{ode1}
\frac{dx}{dt}{}={}u(x,t),
\end{equation}
and the equation for $\ab(t)$ that follows from \eqref{eq:main}:
\begin{equation}
\label{eq:mean}
\frac{d\ab}{dt} {}={} \left((a-c)\ab+(d-b)(1-\ab)\right)\int \frac{s_1s_2}{(s_1+s_2)^3}f(x,t)\,dx.
\end{equation}
The derivation of equations \eqref{eq:main}--\eqref{eq:mean} is contained in the Appendix.

Velocity $u$ represents the rates of change of the stimuli of agents whose current state is given by $x.$  The rates  are proportional to the group average payoffs for corresponding actions, and ``penalized'' by memory for large deviations from the default residual levels $(r_1,r_2).$
Equation \eqref{eq:mean} is convenient for analyzing the games with pure Nash equilibrium. For the game with a mixed equilibrium, described by the frequency $m_*$  to play A, more convenient is equation
\begin{equation}
\label{eq:mean1}
\frac{d(\ab-m_*)}{dt}{}={}
(c-a+d-b)(\ab-m_*)\int \frac{s_1s_2}{(s_1+s_2)^3}f(x,t)\,dx.
\end{equation}

The nonlinear equation \eqref{eq:main} is the first order approximation of the stochastic, learning process. The next order approximation contains a diffusion term, with the diffusion coefficients of order $h.$  The diffusion generally prevents the convergence of learning of the group to a single strategy (fixation).  For example, the convergence of the group learning to a single strategy, based on equation \eqref{eq:main}, only indicates that the  distribution of agents' strategies gets close to that particular strategy, within the limits of validity of equation \eqref{eq:main}. 

Now we consider the dynamics of learning in symmetric $2\times 2$ games. First we consider models with no memory factor, $\mu=0.$

\subsection{Pure Nash equilibrium}
Let $a>c$ and $d>b.$ The characteristic property of this regime is the positive sign of $\ab'(t)$ in equation \eqref{eq:mean}, for any distribution function $f.$ This reflects the fact that  action C is your best choice, no matter what your opponent does.

It is shown in Appendix that $\ab(t)$ increases to its maximum value 1 and the support of function $f$ is transported to infinity along the trajectories of ODE \eqref{ode1}.
In particular the stimuli of all agents, for large values of $t$ will be located below any straight line of positive slope through the origin. That is, asymptotically all agents learn to play the equilibrium strategy ``always A''.
The inclusion the second-order effects does not change the asymptotic picture.

\subsection{Mixed Nash equilibrium}

Let $a<c$ and $d>b.$  The equilibrium mixed strategy is $m_*{}={}\frac{d-b}{c-a+d-b}$ and the group average probability to play A evolves according to equation \eqref{eq:mean1}.

It is shown in Appendix that the dynamics of equation \eqref{eq:main} implies that $\ab(t)$ converges to the equilibrium density $m_*.$ The population average probability to play  A asymptotically coincides with the probability  $m_*$ at the Nash equilibrium. Unlike the pure Nash equilibrium case, in general, agents keep playing with different strategies. This can be seen from the following example. If the initial data $f_0(x)$ describes the population of agents playing different strategies and is such that  
\[
\int \frac{s_1}{s_1+s_2}f_0(x)\,dx{}={}m_*,
\]
then the trajectories of the flow generated by $u$ are straight lines and for any $t>0,$
\[
\int \frac{s_1}{s_1+s_2}f(x,t)\,dx{}={}m_*.
\]
Clearly, for all $t,$ there is non-diminishing spread in the distribution function $f(x,t).$
To put it differently, there is no learning when the group average probability to play A equals $m_*,$ because players expected payoffs from A and B are equal.

\subsection{Two Nash equilibria}
For this type of game  $a>c$ and $d<b.$  Denote by  $m_*{}={}\frac{b-d}{a-c+b-d}.$
As in the previous case, the group average probability to play A evolves according to equation \eqref{eq:mean1}. If the initial data $\ab(0)>m_*$  velocity carries $f$ to 
values of stimulus values of $s_1$ much larger than  $s_2.$ As the average $\ab(t)$ increases,  the dynamics is consistent and leads to learning of the Nash equilibrium A. With $\ab(0)<m_*$ learning converges to the other equilibrium.  In the borderline case $\ab(0)=m_*,$ is unstable: stimuli increase along the straight lines, and agents retain their initial probabilities to play A and B, but any perturbations will deviate the system to A or B equilibrium.

\subsection{Memory factor  RPS}
Consider now models with $\mu>0.$ It can be seen from  equation \eqref{velocity} that  a sufficiently large box $[0,\hat{s}_1]\times[0,\hat{s}_2]$ is invariant under the flow of \eqref{ode1}. This can be seen from the sign of the velocity components. We show in Appendix that with residuals $r_1,r_2>0,$ RPS learning will approach an asymptotically stable point in the stimuli space, for any positive payoff rates $a,b,c,d.$ All agents will tend to play a mixed strategy $m_*{}={}\frac{s^*_1}{s^*_1+s^*_2}.$ When (A,A) is a Nash equilibrium, the ratio of stimuli  $s^*_1/s^*_2>r_1/r_2,$  and agents favor action A after learning more than at their default levels.

\vskip 5pt

There is an interesting limiting case of zero residuals $r_1=r_2=0.$ For such RPS models, when the game has (A,A) as the single Nash equilibrium,  all agents learn this optimal strategy, as function $f(x,t)$ converges to a delta mass supported at the point $(a/\mu,0).$

In this case, the equation for  mean $\ab(t)$ is closely approximated by the replicator dynamics equation \eqref{RDE} given below.  The factor $(s_1(t)+s_2(t))^{-1}$ on the right-hand side of that equation, in this case, is of order 1. The convergence to the optimal strategy is  faster
than in the case of learning with perfect memory ($\mu=0$), for which the factor is of the order $(1+t)^{-1}.$

Moreover, among the models with a memory factor, the learning period appears to be shorter for larger values of $\mu,$ see figure \ref{fig:LT} and explanations in Appendix.
 
  Thus it appears that RPS models with small residuals and large values of $\mu$ should be preferred by natural selection. In such models agents act predominantly  on the basis of the last few payoffs.  In this context it is worth mentioning that  one of the postulates of  prospect theory of Kahneman \& Trverski (1984), is the statement that people actions (in games with monetary payoffs) are directed by the increase in their total wealth, rather than the total accumulated wealth, which shows a tendency to use   short memory and suggests that RPS learning might be at work.

%Taking into account second order effects, equation \eqref{FP},
%that become appreciable when the group moves closer to  $m_*,$
%the process is described by a Ornstein-Uhlenbeck equation with the diffusion coefficient of order $h.$ The distribution settles asymptotically on a 2-d normal distribution with small variance.

\subsection{Relation to the Replicator Dynamics Equation (RDE)}
If one postulates that all agents have the same, or approximately the same, stimuli 
\begin{equation}
\label{H:delta}
X^t_i{}={}(s_1(t),s_2(t)),\quad \forall\, i=1..N,
\end{equation}
for some $s_1(t), s_2(t),$ so that $f(s,t)$ is represented by a delta function supported at $(s_1(t),s_2(t)),$
then equation \eqref{eq:main}, leads to a variant of  the replicator dynamics equation for the probability to cooperate $m(t){}={}s_1(t)/(s_1(t)+s_2(t)):$
\begin{equation}
\label{RDE}
\frac{dm}{dt}{}={}\frac{1}{s_1(t)+s_2(t)}m(1-m)\left( (a-c)m+(d-b)(1-m)\right).
\end{equation}
Notice the  positive factor on the right-hand side of the equation. For a learning processes in which stimuli increase  the learning rate slows down. The extent to which hypothesis \eqref{H:delta} is consistent with the dynamics of \eqref{ode1} is limited only to the cases when the latter has a single asymptotically stable fixed point.

\end{section}

%For later: If the last game was $C,C$ ALL (or some K of agents) (who do not play) agent increase. If $D,D$ all decrease.

%For later: Growing population

%\begin{section}{Summary}
%We analyzed the RPS learning rule for generic values of  parameters, using the mean field approximation of the stochastic learning process with infinitesimal increments. It was established that rules with memory decay converge to the equilibrium of a 2x2 symmetric game faster than rules with ``perfect memory''.  For the rules with memory decay the characteristic time of convergence is of the order $\mu^{-1}.$ 

%Thus none of them is an evolutionarily stable (ES) learning rule, in the sense of Harley (1981). 

%The RPS model with zero memory, ($\bar{\mu}=0$ in \eqref{RPS1}, \eqref{RPS2}), on the other hand, will not let organisms discover the optimal behavior. It appears that an ES learning rule must have some memory, but shorter than the memory of the RSP rules. 

%\end{section}

\begin{section}{Appendix: a PDE model}
\subsection{Fokker-Planck equation}
Consider a group of $N$ individuals acting according to RPS learning rule described in section \ref{sec1}. Let $X^t=(X^t_1,..,X^t_N)$ represent the vector of pairs of stimuli for all members at epoch $t.$ Each component of this vector is 2-dimensional:  $X^t_i{}={}(S^t_{1,i},S^t_{2,i}).$  By $w_h(\xb,t),$ where $x\in[0,1]^{2N},$ we denote PDF for distribution of $X^t.$ We will write $\xb=(x_1,..x_N),$ where each $x_i{}={}(s_{1,i},s_{2,i}).$ The probability to play A will be denoted as $\lambda_i{}={}s_{1,i}/(s_{1,i}+s_{2,i}).$

 Suppose that member $i$ and $j$ are selected for the interaction. There will be only one game played during the period from $t$ to $t+\delta.$ The matrix of payoffs is described in table \ref{table:inc_pay}. The range of parameters $\delta,\,h,\,N$ will be restricted later on.

Conditioned on the event $X^t=\xb,$ the agent probabilities for the next period are set according to the RPS rule \eqref{eq:RPS}, which in the notation of the stochastic process are
\[
X^{t+\delta}_i{}={}\left\{
\begin{array}{ll}
((1-\mu h)s_{1,i} + r_1\mu h +ah,\, (1-\mu h)s_{2,i}+r_2\mu h) & \rm Prob=\lambda_i\lambda_j\\
((1-\mu h)s_{1,i} +r_1\mu h  + dh,\,(1-\mu h)s_{2,i}+r_2\mu h) & \rm Prob=\lambda_i(1-\lambda_j)\\
((1-\mu h)s_{1,i} +r_1\mu h,\,(1-\mu h)s_{2,i}+r_2\mu h + ch) & \rm Prob=(1-\lambda_i)\lambda_j\\
((1-\mu h)s_{1,i} +r_1\mu h,\,(1-\mu h)s_{2,i} +r_2\mu h + bh) & \rm Prob=(1-\lambda_i)(1-\lambda_j)
\end{array}
\right.
\]
and symmetrically for $X^{t+\delta}_j.$
For all other agents, $X^{t+\delta}_k{}={}X^t_k$ for $k\not=i,j.$ The definition of $X^t$ makes it a discrete-time Markov process. We proceed by writing down the integral form of the Chapman-Kolmogorov equations and approximate its solution by a solution of the Fokker-Planck equation (forward Kolmogorov's equation), for small values of $\delta, h$ and large $N.$ 

Change of $w_h(\xb,t)$ from $t$ to $t+\delta,$ can be described in the following way. 
\begin{multline}
\int \phi(\xb)w_h(\xb,t+\delta)\,d\xb{}={}\mathbb{E}[\phi(X^{t+h})]\\
{}={}\sum_{i\not=j}(N(N-1))^{-1}\int\left(
\lambda_i\lambda_j\phi(\xb)\Big|_{x_i{}={}((1-\mu h)s_{1,i} + r_1\mu h +ah,\, (1-\mu h)s_{2,i}+r_2\mu h)\atop 
\,\,\, x_j=((1-\mu h)s_{1,j} + r_1\mu h +ah,\, (1-\mu h)s_{2,j}+r_2\mu h)}\right.\\
\left.
+ \lambda_i(1-\lambda_j)\phi(\xb)\Big|_{x_i{}={}((1-\mu h)s_{1,i} +r_1\mu h  + dh,\,(1-\mu h)s_{2,i}+r_2\mu h)\atop
\,\,\, x_j{}={}((1-\mu h)s_{1,j} +r_1\mu h,\,(1-\mu h)s_{2,j}+r_2\mu h + ch)}\right. \\
\left. 
+(1-\lambda_i)\lambda_j\phi(\xb)\Big|_{x_i=((1-\mu h)s_{1,i} +r_1\mu h,\,(1-\mu h)s_{2,i}+r_2\mu h + ch) \atop  \, \,\, x_j=((1-\mu h)s_{1,j} +r_1\mu h  + dh,\,(1-\mu h)s_{2,j}+r_2\mu h)} \right. \\
\left. 
+(1-\lambda_i)(1-\lambda_j)\phi(\xb)\Big|_{x_i=((1-\mu h)s_{1,i} +r_1\mu h,\,(1-\mu h)s_{2,i} +r_2\mu h + bh) \atop \,\,\, x_j=((1-\mu h)s_{1,j} +r_1\mu h,\,(1-\mu h)s_{2,j} +r_2\mu h + bh)}\right)w_h(\xb,t)\,d\xb.
\end{multline}
This equation can be written in slightly different way:
\begin{multline}
\label{eq:int1}
\int \phi(\xb)w_h(\xb,t+\delta)\,d\xb
{}={}\int \phi(\xb)w_h(x,t)\,d\xb \\
{}+{}\sum_{i\not=j}(N(N-1))^{-1}\int\left(
\lambda_i\lambda_j[\phi(\xb)\Big|_{x_i{}={}((1-\mu h)s_{1,i} + r_1\mu h +ah,\, (1-\mu h)s_{2,i}+r_2\mu h)\atop 
\,\,\, x_j=((1-\mu h)s_{1,j} + r_1\mu h +ah,\, (1-\mu h)s_{2,j}+r_2\mu h)}-\phi(\xb)]\right.\\
\left.
+ \lambda_i(1-\lambda_j)[\phi(\xb)\Big|_{x_i{}={}((1-\mu h)s_{1,i} +r_1\mu h  + dh,\,(1-\mu h)s_{2,i}+r_2\mu h)\atop
\,\,\, x_j{}={}((1-\mu h)s_{1,j} +r_1\mu h,\,(1-\mu h)s_{2,j}+r_2\mu h + ch)}-\phi(\xb)]\right. \\
\left. 
+(1-\lambda_i)\lambda_j[\phi(\xb)\Big|_{x_i=((1-\mu h)s_{1,i} +r_1\mu h,\,(1-\mu h)s_{2,i}+r_2\mu h + ch) \atop  \, \,\, x_j=((1-\mu h)s_{1,j} +r_1\mu h  + dh,\,(1-\mu h)s_{2,j}+r_2\mu h)}-\phi(\xb)] \right. \\
\left. 
+(1-\lambda_i)(1-\lambda_j)[\phi(\xb)\Big|_{x_i=((1-\mu h)s_{1,i} +r_1\mu h,\,(1-\mu h)s_{2,i} +r_2\mu h + bh) \atop \,\,\, x_j=((1-\mu h)s_{1,j} +r_1\mu h,\,(1-\mu h)s_{2,j} +r_2\mu h + bh)}-\phi(\xb)]\right)w_h(\xb,t)\,d\xb.
\end{multline}

\vskip 10pt

The above equation can be used to obtain $2N$-dimensional ODE approximation of the stochastic process by  evaluating $\lim_{h\to0}\left(\mathbb{E}[X^{t+h}\,|\, X^t]-X^t\right)/h{}={}F(X^t).$ This approach was implemented in the method of stochastic approximation developed by Benaim-Hirsch (1999) and applied to the study of convergence of stochastic fictitious play processes. The method guarantees the convergence of the process $X^t$ under certain stability conditions for the dynamics of the associated ODE. 

The large dimension of that dynamical system is an obstacle for further analysis. In contrast, we would like to obtain an equation for the distribution of large number $N$ of agents in 2-dimensional stimuli space.  For this, denote the PDF of the distribution by 
\[
f_h(x,t){}={}\sum_k N^{-1}\int w_h(\xb)\big|_{x_k=x}\,d\xb_k,\quad x\in\mathbb{R}^2,
\]
where $\xb_k$ is a $2N-2$ dimensional vector of all coordinates, excluding $x_k.$
In statistical physics this function is also called one-particle distribution. In the formulas to follow we need to use two-particle distribution function 
\[
g_h(x,y,t){}={}\sum_{i\not=j} (N(N-1))^{-1}\int w_h(\xb)\big|_{x_i=x,\, \\x_j=y}\,d\xb_{ij},
\]
where $\xb_{ij}$ is the $2N-4$ dimensional vector of all coordinated excluding $x_i$ and $x_j.$ Function $g_h$ is symmetric in $(x,y)$ and is related to $f_h$ by the formulas
\[
f_h(x,t){}={}\int g_h(x,y,t)\,dx{}={}\int g_h(x,y,t)\,dy.
\]
The moments of function $f_h$ and $g_h$ are computed from the  moment of $w_h:$
\[
\int \psi(x)f_h(x,t)\,dx{}={}\sum_k N^{-1}\int \psi(x_k)w_h(\xb)\,d\xb,
\]
and 
\[
\int \omega(x,y)g_h(x,y,t)\,dxdy{}={}\sum_{i\not=j} (N(N-1))^{-1}\int \omega(x_i,x_j)w_h(\xb)\,d\xb.
\]
This follows from the definition of these functions.

Now we use \eqref{eq:int1} to obtain an integral equation of the change of function $f_h.$ For that select $\phi(\xb){}={}\psi(x_k),$ sum over $k$ and take average. We get
\begin{multline}
\int \psi(x)f_h(x,t+\delta)\,dx{}={}\int \psi(x)f_h(x,t)\,dx\\
{}+{}N^{-1}\sum_{i\not=j}(N(N-1))^{-1}\int\left(
\lambda_i\lambda_j[\psi((1-\mu h)s_{1,i} + r_1\mu h +ah,\, (1-\mu h)s_{2,i}+r_2\mu h) - \psi(x_i)  \right.
\\ \left.
+\psi((1-\mu h)s_{1,j} + r_1\mu h +ah,\, (1-\mu h)s_{2,j}+r_2\mu h)-\psi(x_j)]\right. \\
\left. \right.\\ \left.
+\lambda_i(1-\lambda_j)[\psi((1-\mu h)s_{1,i} +r_1\mu h  + dh,\,(1-\mu h)s_{2,i}+r_2\mu h)-\psi(x_i)\right. \\
\left. + \psi((1-\mu h)s_{1,j} +r_1\mu h,\,(1-\mu h)s_{2,j}+r_2\mu h + ch) - \psi(x_j)] \right. \\
\left. \right. \\
\left. +(1-\lambda_i)\lambda_j[ \psi((1-\mu h)s_{1,i} +r_1\mu h,\,(1-\mu h)s_{2,i}+r_2\mu h + ch)-\psi(x_i) \right. \\
\left. + \psi((1-\mu h)s_{1,j} +r_1\mu h  + dh,\,(1-\mu h)s_{2,j}+r_2\mu h) - \psi(x_j)] \right. \\
\left. \right. \\ \left. +(1-\lambda_i)(1-\lambda_j)[\psi((1-\mu h)s_{1,i} +r_1\mu h,\,(1-\mu h)s_{2,i} +r_2\mu h + bh)-\psi(x_i)   \right. \\
\left. + \psi((1-\mu h)s_{1,j} +r_1\mu h,\,(1-\mu h)s_{2,j} +r_2\mu h + bh) - \psi(x_j)] \right)w_h(\xb,t)\,d\xb.
\end{multline}
The right-hand side can be conveniently expressed in terms of the two-particle function $g_h:$

\begin{multline}
\label{int_f}
\int \psi(x)f_h(x,t+\delta)\,dx{}={}\int \psi(x)f_h(x,t)\,dx\\
{}+{}2N^{-1}\int\left(
\lambda(x)\lambda(y)[\psi((1-\mu h)s^x_1 + r_1\mu h +ah,\, (1-\mu h)s^x_2+r_2\mu h) - \psi(x)  \right.
\\ \left.
+\psi((1-\mu h)s^y_1 + r_1\mu h +ah,\, (1-\mu h)s^y_2+r_2\mu h)-\psi(y)]\right. \\
\left. \right.\\ \left.
+\lambda(x)(1-\lambda(y))[\psi((1-\mu h)s^x_1 +r_1\mu h  + dh,\,(1-\mu h)s^x_2+r_2\mu h)-\psi(x)\right. \\
\left. + \psi((1-\mu h)s^y_1 +r_1\mu h,\,(1-\mu h)s^y_2+r_2\mu h + ch) - \psi(y)] \right. \\
\left. \right. \\
\left. +(1-\lambda(x))\lambda(y)[ \psi((1-\mu h)s^x_1 +r_1\mu h,\,(1-\mu h)s^x_2+r_2\mu h + ch)-\psi(x) \right. \\
\left. + \psi((1-\mu h)s^y_1 +r_1\mu h  + dh,\,(1-\mu h)s^y_2+r_2\mu h) - \psi(y)] \right. \\
\left. \right. \\ \left. +(1-\lambda(x))(1-\lambda(y))[\psi((1-\mu h)s^x_1 +r_1\mu h,\,(1-\mu h)s^x_2 +r_2\mu h + bh)-\psi(x)   \right. \\
\left. + \psi((1-\mu h)s^y_1 +r_1\mu h,\,(1-\mu h)s^y_2 +r_2\mu h + bh) - \psi(y)]
\right)g_h(x,y,t)\,dxdy.
\end{multline}
where $x=(s^x_1,s^x_2),$ $y=(s^y_1,s^y_2),$ and $\lambda(x){}={}s^x_1/(s^x_1+s^x_2),$ and similar for $\lambda(y).$
In the processes with large number of agents and random binary interactions, two-particle distribution function can be factored into two independent distributions:
\[
g_h(x,y,t){}={}f_h(x,t)f_h(y,t).
\]
With this relation, \eqref{int_f}, becomes a family of non-linear integral relations for the next time step distribution $f_h(x,t+\delta).$ 
Taking the Taylor expansions up to the first order for the increment of the test function $\psi,$ we obtain integral equations:
\begin{multline*}
\frac{N}{2h}\int \psi(x)(f_h(x,t+\delta)-f_h(x,t))\,dx{}={}\\
 2\ab_h(t)\int\left(
\lambda(x)(a + \mu(r_1-s_1))\partial_{s_1}\psi(x)
{}+{}
\lambda(x)\mu(r_2-s_2)\partial_{s_2}\psi(x)\right)
f_h(x,t)\,dx \\
{}+{}2(1-\ab_h(t))\int\left(
\lambda(x)(d + \mu(r_1-s_1))\partial_{s_1}\psi(x)
{}+{}
\lambda(x)\mu(r_2-s_2)\partial_{s_2}\psi(x)\right)
f_h(x,t)\,dx \\
{}+{}2\ab_h(t)\int\left(
(1-\lambda(x))(\mu(r_1-s_1))\partial_{s_1}\psi(x) \right.\\
\left. {}+{}
(1-\lambda(x))(c+\mu(r_2-s_2))\partial_{s_2}\psi(x)\right)
f_h(x,t)\,dx \\
{}+{}2(1-\ab_h(t))\int\left(
(1-\lambda(x))(\mu(r_1-s_1))\partial_{s_1}\psi(x)\right.
\\ \left. {}+{}
(1-\lambda(x))(b+\mu(r_2-s_2))\partial_{s_2}\psi(x)\right)
f_h(x,t)\,dx
\end{multline*}
where $\ab_h(t){}={}\int \lambda(x)f_h(x,t)\,dx.$

Combining various terms on the right-hand side of the equation we get
\[
\frac{N}{4h}\int \psi(x)(f_h(x,t+\delta)-f_h(x,t))\,dx{}={}\\
\int \left(v_1\partial_{s_1}\psi +v_2\partial_{s_2}\psi\right)f_h(x,t)\,dx,
\]
with 
\[
v_1{}={}\lambda(x)(a\ab_h(t)+ (1-\ab_h(t))d) + \mu(r_1-s_1),
\]
\[
v_2{}={}(1-\lambda(x))(c\ab_h(t)+ (1-\ab_h(t))b) + \mu(r_2-s_2).
\]

By integrating the right-hand side by parts, and assuming that  $h,\delta$ are small and $N$ is large, in such a way that  $\frac{4h}{\delta N}= 1,$ we obtain the Fokker-Planck equation:
\begin{equation}
\label{FP}
\frac{\partial f}{\partial t}{}+{}\div (u(x,t)f){}={}0,
\end{equation}
where $x=(s_1,s_2),\,s_1,s_2>0$ and the drift velocity is given by the formula
\[
u(x,t){}={}\frac{1}{s_1+s_2}\left[
\begin{array}{l}
\left(a\ab(t){}+{}d(1-\ab(t)\right)s_1 + \mu(r_1-s_1) ( s_1+s_2) \\
\left( c\ab(t){}+{}b(1-\ab(t)\right)s_2 +\mu(r_2-s_2)  (s_1 +s_2)
\end{array}
\right],
\]
where 
\begin{equation}
\label{m}
\ab(t){}={}\int\frac{s_1}{s_1+s_2}f(x,t)\,dx.
\end{equation}

Equation \eqref{eq:mean} is obtained from \eqref{FP} by multiplying it by $\lambda(x){}={}s_1(s_1+s_2),$ and integrating by parts. We're assuming here that the support of $f$ is contained in the interior of the first quadrant, so that $f$ is zero on the boundary. That is to say that all agents play mixed strategies. This is a natural hypothesis, as nothing else can be learned if an agent chooses an action with certainty.

Consider now the learning  from playing a symmetric $n\times n$ game. Let the payoff to playing $i^{th}$ action against $j^{th}$ be equal to $ a_{ij}h.$ Denote the stimulus vector $x=(s_1,..,s_n),$ and the population average probability to play $i$ by
\[
\ab_i(t){}={}\int \frac{s_i}{\sum_j s_j}f(x,t)\,dx,\quad i=1..n.
\]
The first-order approximation of the RPS learning process is given by the Fokker-Planck equation
\[
\frac{\partial f}{\partial t} {}+{}\div (u(x,t)f){}={}0,
\]
on the domain with $s_i\geq0,$ $i=1..n.$
In this equation,  the velocity vector $u=(u_1,..,u_n)$ is given by its components
\[
u_i(x,t){}={}\frac{1}{\sum_j s_j}\left[
\left(\sum_j a_{ij}\ab_j(t)\right)s_i + \mu(r_i-s_i) \sum_j s_j\right],\quad i=1..n.
\]

\vskip 5pt

Now we consider in some detail the learning in 2x2 games. Much of the analysis of equation \eqref{FP} is derived from the behavior of trajectories of  ODE \eqref{ode1}.
The solution $f$ of \eqref{FP} is obtained by transporting the support of $f_0$ along trajectories of the dynamical system \eqref{ode1} and changing the values  $f_0$ so that the ``mass'' (measured by the density function  $f$) of any ``fluid element" remains constant. In fact on can write down the formula for $f$ in terms of $u$ and prove that the solution $f$ of the non-linear problem exists and unique. This can be done  by standard methods of PDEs, but it is outside of the scope of the present paper.  Here, we will be interested  in the long time, qualitative asymptotic for   $f(x,t).$

Equation \eqref{FP} is considered in the first quadrant $s_1,s_2>0.$ For $\mu=0,$ the boundary of the domain in invariant under the flow of \eqref{ode1}. For the model with fading memory, $\mu>0,$ the velocity $u$ at the boundary  is directed into the flow domain.  In either case, we will assume that the function $f_0(x)$ is zero on the boundary. Then, this property will hold for all $t>0.$ Additionally, in all of the analysis below, we assume that $f_0$ is a continuously differentiable function with compact support (zero outside some bounded set). 

Consider the case of the pure Nash equilibrium ($a>c,d>b$) and no memory effects, $\mu=0.$
The velocity 
\[
u(x,t){}={}\frac{1}{s_1+s_2}M(t)x,
\]
where matrix
\[
M(t){}={}\left[
\begin{array}{cc}
a\ab(t){}+{}d(1-\ab(t)) & 0\\
0 & c\ab(t){}+{}b(1-\ab(t))
\end{array}
\right]
\]
For any $t,$ the origin is  an unstable node with two positive eigenvalues; the eigenvalue corresponding to $s_1$--direction is the dominant one. From the phase portrait of the ODE it is clear that the flow transports the support of $f$ into the region where $s_1\gg s_2,$ which correspond to the case of all agents asymptotically in $t$ adopting choice C  in the game.

In the case of the mixed Nash equilibrium ($a<c,d>b$) and no memory effect, $\mu=0,$
the origin is  an unstable node. When  $\ab(t)=m_*$ then  two positive eigenvalues coincide, and all trajectories are straight lines through the origin. In general, however, $\ab(t)\not=m_*,$ if for example they are not equal at time $t=0.$ In such cases equation \eqref{eq:mean1} can be used to show that $\lim_{t\to\infty}\ab(t)=m_*. $ Let $\ab(0) > m_*.$ Then, according to \eqref{eq:mean1}, $\ab(0)>\ab(t)>m_*$ for all $t,$ and $\ab(t)$ converges to $m_*$ provided that  
\begin{equation}
\label{diverge}
\int_0^{\infty}\int \frac{s_1s_2}{(s_1+s_2)^{3}}f(x,t)\,dx dt
\end{equation}
 diverges. Notice also, that  the derivative of ratio $s_2/s_1$ along a flow trajectory equals
\[
\frac{d}{dt}\left(\frac{s_2}{s_1}\right) {}={}(c-a+d-b)(\ab(t)-m_*)\frac{s_1s_2}{s_1+s_2}>0.
\]
Thus, if at $t=0$ the support of $f_0(\cdot)$ is strictly inside the first quadrant, then there $c>0$ such that $s_2/s_1>c$ for all points in the support of $f(\cdot,t)$ for all later times. In particular, one can estimate
\begin{multline}
\label{ineq:converge}
\int\frac{s_1s_2}{(s_1+s_2)^3}f(x,t)\,dx{}={}\int \frac{s_2/s_1}{(1 + s_2/s_1)(s_1+s_2)}\frac{s_1}{s_1+s_2}f(x,t)\,dx\\
>\sup_{(s_1,s_2)\in {\rm supp} f(\cdot,t)}\frac{c}{(1+c)(s_1+s_2)}\ab(t).
\end{multline}
Finally,since $u(x,t)$ is uniformly bounded, i.e., for any $(x,t),$ $|u(x,t)|< C,$ for some $C,$ then for any $x$ in the support of  $f(\cdot,t)$ there a constant $\hat{C}$ such that $|x|<\hat{C}(1+t).$ From this and \eqref{ineq:converge} it follows that
\[
\int \frac{s_1s_2}{(s_1+s_2)^3}f(x,t)\,dx>Cm_*(1+t)^{-1},
\]
for some constant $C>0,$ and so, the integral in \eqref{diverge} is infinite. The case $\ab(0)<m_*$ follows by similar arguments.

Consider now the model with the memory decay when  $\mu>0$ and residuals $r_1,r_2>0.$ For any value of $\ab(t)\in[0,1]$ and any set of positive parameters of the game $a,b,c,d>0$ the right-hand side of \eqref{ode1} has a steady state $(s^0_1(t),s^0_2(t))$  in the interior of the first quadrant, with $s^0_1>r_1,$ $s^0_2>r_2,$  and this point is an asymptotically  stable node. The other steady state is the origin, which is an unstable node. The support of $f(x,t)$ moves in the direction of the stable interior point, contracting in size. When it becomes sufficiently small, the dynamics can be effectively approximated by ODE: $\frac{d(s_1,s_2)}{dt}{}={}(\bar{u}_1,\,\bar{u}_2)$ where the new velocity
\begin{eqnarray*}
\label{u1}
\bar{u}_1=\frac{a(s_1)^2}{(s_1+s_2)^2}{}+{}\frac{ds_1s_2}{(s_1+s_2)^2}{}+{}\mu(r_1-s_1),\\
\label{u2}
\bar{u}_2=\frac{cs_1s_2}{(s_1+s_2)^2}{}+{}\frac{b(s_2)^2}{(s_1+s_2)^2}{}+{}\mu(r_2-s_2).
\end{eqnarray*}
In a long run the fixed point will settle at the stable, interior state  $x^*=(s^*_1,s^*_2),$ and $f(x,t)$ will be a delta-function supported at that point.  In this process the agents learn  to play A with probability $m_*{}={}s^*_1/(s^*_1+s^*_2).$

A special case of zero residuals $r_1=r_2=0$ deserves a discussion. In the limit of zero residuals $r_1,r_2\to0,$ velocity $u$ (for any fixed $t>0$) has three fixed points:
$x_0(t)=((a\ab(t)+d(1-\ab(t))/\mu,0),$ $x_1=(0,(c\ab(t)+b(1-\ab(t))/\mu)$ and $(0,0).$ When $a>c,d>b,$ the first, corresponding to the strategy ``always A'', is an asymptotically stable node, the second, corresponding to ``always B'', is a saddle, and the origin is an unstable node. One can compute that  on any trajectory of the velocity field $u(x,t)$ inside the first quadrant,
\[
\frac{d}{dt}\left(\frac{s_1}{s_2}\right)>0.
\]
Thus, the population average probability to play A, $\ab(t),$ increases to 1, the stable stationary point $x_0(t)$ converges to $x_0=(a/mu,0),$ and the support of $f(x,t)$ moves toward point $x_0.$  The agents with  memory decay and zero residual levels do learn the optimal strategy. Moreover, because the learning occurs in the bounded region of the stimuli space, the convergence to the equilibrium is faster than the case  of learning with perfect memory $\mu=0.$ Using equation \eqref{RDE} we can also estimate on the rate of convergence as a function of $\mu.$ The rate is proportional to $\mu$ implying that the characteristic time of convergence to be $C/\mu.$ Figure  \ref{fig:LT} shows the simulations  of the stochastic process in this regime for different values of $\mu.$ It shows that the prediction based on the model \eqref{FP} is in good agreement with the stochastic learning process.
We conclude that models with low residuals and high $\mu$ perform better in learning the optimal strategy in 2x2 games and thus more likely to evolve.

%\begin{figure}[t]
%\centering
%\includegraphics[clip, trim=1.5cm 7.5cm 1.5cm 3.2cm, width=7cm]{PPlane_JTB}
%\caption{The directional field and several trajectories of \eqref{FP} at time $t=0$ for RPS learning with $\mu=1,$ $r_1=r_2=0,$ and  a=3, b=1, c=d=2. At $t>0$ direction field has similar structure but the fixed points on x($=s_1$) and y($=s_2$) axis change their locations. The figure
%was produces by pplane software by John C. Polking, %Department of Mathematics, Rice University. \label{fig:PP}}
%\end{figure}

\begin{figure}[t]
\centering
\includegraphics[width=8cm]{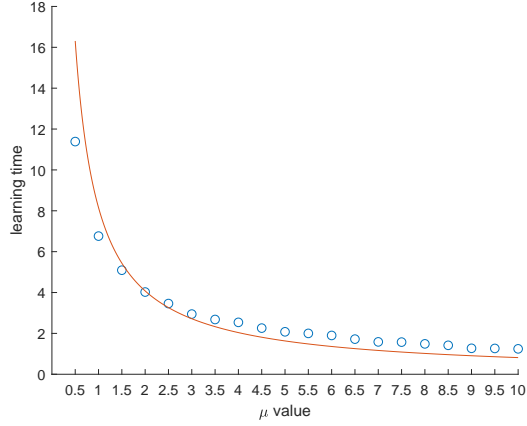}
\caption{Learning time to equilibrium in the game with a=3, b=1, c=d=2, h=0.01, N=100 and $\mu$ ranging from 0.5 to 10 with increment of 0.5. All agents have stimuli (1,1) at the beginning of the process. Learning time is a time it takes for at least 90\% of population to have probabilities to play A in the range [0.9,1]. Each learning time is the average over 20 simulations. Units of time are arbitrary. The line plot is the best fit to the data points by  function $c_0\mu^{-1}.$ The figure was generated using MatLab R2017b software. \label{fig:LT}}
\end{figure}

\subsection{Second order effects}
The inclusion of the next order approximation into consideration adds a diffusion term into equation \eqref{FP} with the diffusion coefficients proportional to $h.$ In the problems where the drift velocity takes $f$ to regions with large values of $x,$ as in the case of the pure or mixed Nash equilibrium, diffusion will have a marginal effect. In the problems with asymptotically stable fixed points in stimuli space (short memory models) diffusion will create a stationary distribution
of $f$ near the fixed point, preventing all agents to adopt a single strategy.

\end{section}

The author wishes to thank the anonymous referees for patient reading of the manuscript and detailed comments that helped to improve it in so many ways.

\end{document}